\documentclass[superscriptaddress,showpacs,preprintnumbers,amsmath,amssymb]{revtex4}

\usepackage{graphicx}% Include figure files
\usepackage{dcolumn}% Align table columns on decimal point
\usepackage{bm}% bold math

\begin{document}

\preprint{}

\title{Exploring Hadron Physics in Black Hole Formations: \\ a New Promising Target of Neutrino Astronomy}% Force line breaks with \\

\author{Ken'ichiro Nakazato}
% \altaffiliation[Present address: ]{Department of Astronomy, Kyoto University, Kita-shirakawa Oiwake-cho, Sakyo, Kyoto 606-8502, Japan
%, 3-4-1 Okubo, Shinjuku, Tokyo, Japan
%}
 \email{nakazato@kusastro.kyoto-u.ac.jp}
 \affiliation{Department of Astronomy, Kyoto University, Kita-shirakawa Oiwake-cho, Sakyo, Kyoto 606-8502, Japan
}%

\author{Kohsuke Sumiyoshi}
\affiliation{Numazu College of Technology, Ooka 3600, Numazu, Shizuoka 410-8501, Japan
}%

\author{Hideyuki Suzuki}
\affiliation{Faculty of Science and Technology, Tokyo University of Science, Yamazaki 2641, Noda, Chiba 278-8510, Japan
}%

\author{Shoichi Yamada}%
 \altaffiliation[Also at ]{Advanced Research Institute for Science and Engineering, Waseda University, 3-4-1 Okubo, Shinjuku, Tokyo 169-8555, Japan
}
 \affiliation{Department of Physics, Waseda University, 3-4-1 Okubo, Shinjuku, Tokyo 169-8555, Japan
}%

\date{\today}% It is always \today, today,

\begin{abstract}
The detection of neutrinos from massive stellar collapses can teach us a lot not only about source objects but also about microphysics working deep inside them. In this study we discuss quantitatively the possibility to extract information on the properties of dense and hot hadronic matter from neutrino signals coming out of black-hole-forming collapses of non-rotational massive stars. Based on our detailed numerical simulations we evaluate the event numbers for SuperKamiokande with neutrino oscillations being fully taken into account. We demonstrate that the event numbers from a Galactic event are large enough not only to detect it but also to distinguish one hadronic equation of state from another by our statistical method assuming the same progenitor model and non-rotation. This means that the massive stellar collapse can be a unique probe into hadron physics and will be a promising target of the nascent neutrino astronomy.
\end{abstract}

%An article usually includes an abstract, a concise summary of the work
%covered at length in the main body of the article. It is used for
%secondary publications and for information retrieval purposes. Valid
%PACS numbers may be entered using the \verb+\pacs{#1}+ command.
\pacs{26.50.+x, 95.85.Ry, 97.60.-s, 21.65.Mn}% PACS, the Physics and Astronomy
%\keywords{}%Use showkeys class option if keyword
%                              %display desired

\maketitle

\section{Introduction} \label{intro}

One of the important roles of astrophysics is to explore physics under extreme conditions that are difficult to realize in terrestrial experiments. In this sense, hadron physics at supra-nuclear densities and non-zero temperatures (e.g., hyperon appearance, quark deconfinement and so on) is a natural target of astrophysics and the gravitational collapse of massive stars at the end of their lives will set the stage \cite{sumi09, self08a}. In particular, black-hole-forming collapses expected for very massive stars with a mass larger than $\sim \!\! 30$ solar masses ($M_\odot$) will be the most promising site. Although such an event has not been observed yet, a black hole candidate with an estimated mass of 24-33$M_{\odot}$ was discovered \cite{prest07} and this might be a remnant of the collapse of such a massive star. Recently, a regular monitoring of $\sim \!\! 10^6$ supergiants within a distance of 10~Mpc is proposed to detect their silent disappearances \cite{kocha08} and the black-hole-forming collapse, which would be invisible optically, might be observed that way. 

In our previous studies \cite{sumi06, sumi07, sumi08}, we showed that the event, which we refer to as the ``failed supernova'' hereafter, is as bright in neutrino emissions as ordinary core-collapse supernovae. We showed also that its time evolutions of luminosities and spectra are qualitatively different from those of the ordinary supernova explosion and  the ensuing proto-neutron star cooling \cite{totani98}, which may lead to the delayed black hole formation for some reasons \cite{refb, refb2}. Our numerical data were adopted as a reliable basis to predict the relic neutrino background from stellar collapses \cite{luna09}. More importantly, however, we also demonstrated by employing different hadronic equations of state (EOS) that the duration of neutrino emissions from the failed supernova is sensitive to the stiffness of EOS at supra-nuclear densities and, therefore, that the observation of neutrinos from such an event will provide us with valuable information on the properties of dense and hot hadronic matter as well as on the maximum mass of proto-neutron stars.

Although this approach is simple and robust, valid irrespective of neutrino oscillations, it can not distinguish EOS's with a similar duration of neutrino emissions: a soft nucleonic EOS and a hyperonic EOS, for example. In this study, we attempt to break this degeneracy by analyzing more in detail the time variation of neutrino numbers observed at a terrestrial detector, which we refer to as the ``light curve'' hereafter. While we have studied the detection of failed supernova neutrinos taking fully into account the neutrino oscillation and its parameter dependence so far \cite{self08b}, we innovate a new method here by employing the Kolmogolov-Smirnov (KS) test, which is free from the ambiguity of the distance to the progenitor. We adopt the results of our detailed numerical simulations and evaluate the neutrino event number for SuperKamiokande (SK) as a representative of currently operating neutrino detectors. This is the first ever serious self-contained attempt to demonstrate that for Galactic events it is indeed possible to break the degeneracy for hadronic EOS's by the statistical analysis.

We arrange this paper as follows. A brief review of the neutrino detection and a description of the newly proposed statistical method are given in Sec.~\ref{setup}. The main results of our study are reported in Sec.~\ref{result}. In Sec.~\ref{disc}, we mention the possible uncertainties and observational issues. Sec.~\ref{sum} is devoted to a summary.

\section{Methods} \label{setup}

\subsection{Setups for failed supernova neutrino detection}

The evaluation of the light curve of neutrinos from the failed supernova can be roughly divided into three parts. The first step is a computation of the neutrino luminosity and spectrum at the source. The general relativistic $\nu$-radiation-hydrodynamics code, which solves the Boltzmann equations for neutrinos together with the Lagrangian hydrodynamics under spherical symmetry, is utilized to quantitatively compute the dynamics as well as the neutrino luminosities and spectra up to the black hole formation. This code passed a couple of well known standard tests and a detailed comparison with Monte Carlo simulations \cite{yamada97, yamada99, sumi05}. The numerical errors are estimated to be $\sim \!\! 10$\% from the computations with lower resolutions in Ref.~\cite{self07}. The progenitor model with $40M_{\odot}$ \cite{woosley95} is adopted as the initial condition for the dynamical simulations.

A hadronic EOS is needed at this stage. It should be emphasized that it is not the intention of the paper to endorse a particular EOS but that EOS's which are available for astrophysical numerical simulations, that is, subroutines or tables that provide thermodynamic variables in wide ranges of density, temperature and proton fraction are very limited at present. For example, an EOS table including hyperons has been provided only by Ishizuka et~al. \cite{takochu} so far, based on the relativistic mean field theory.  The EOS's by Lattimer \& Swesty \cite{lati91} and by Shen et~al. \cite{shen98a, shen98b} are the ones from the limited options for the nucleonic EOS. The former is an extension of the compressible liquid drop model with three choices of incompressibility ($K=180$, 220, 375~MeV) and the latter is based on the same frame work as in Ref.~\cite{takochu} but without hyperons. In this paper, we employ the EOS's by Ref.~\cite{takochu} (Hyperon-EOS), Ref.~\cite{shen98a, shen98b} (Shen-EOS) and Ref.~\cite{lati91} with $K=180$~MeV (LS180-EOS) and 220~MeV (LS220-EOS). Whereas the results for Hyperon-EOS, Shen-EOS and LS180-EOS have already been given also in Ref.~\cite{sumi09}, LS220-EOS model is newly computed.

\begin{figure}
\begin{center}
\includegraphics{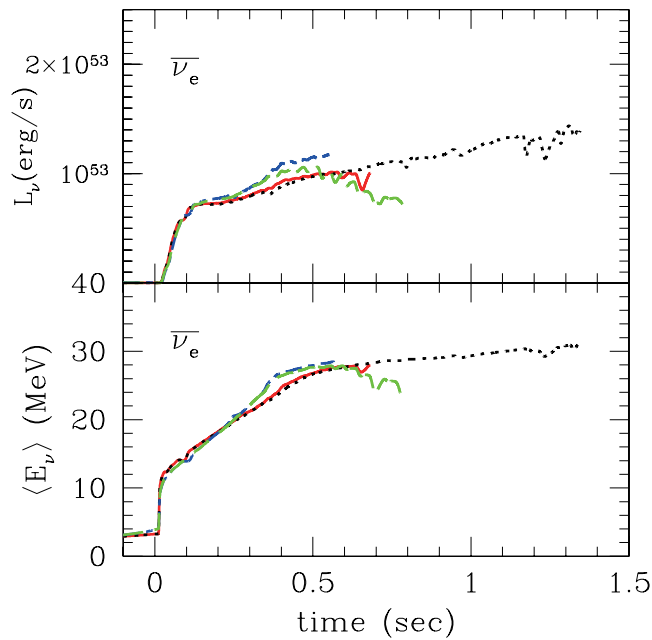}
\caption{Luminosities and average energies of $\bar\nu_e$ for four different EOS's, LS180-EOS (dash-dotted), LS220-EOS (dashed), Shen-EOS (dotted) and Hyperon-EOS (solid). Time is measured from the bounce.}
\label{sumires}
\end{center}
\end{figure}
In Fig.~\ref{sumires}, we show the luminosities and average energies of electron-type anti-neutrino for these EOS's. The result for Hyperon-EOS is almost the same as that for Shen-EOS until hyperons appear around $\sim \!\! 0.5$~sec, since the two EOS's are identical in the low density regime. The Shen-EOS, which is the stiffest ($K=281$~MeV) among these EOS's, is sufficiently distinguishable from the other three just by its longer duration (1.345~sec) of neutrino emissions. This will not be the case, on the other hand, for Hyperon-EOS and LS-EOS's with $K\sim200$~MeV because the duration for Hyperon-EOS (0.682~sec) is not very different from those for LS180-EOS (0.566~sec) and LS220-EOS (0.784~sec). This is the degeneracy problem mentioned earlier. It is the main purpose of the study to demonstrate that the relatively small difference displayed in Fig.~\ref{sumires} is still sufficient to distinguish one EOS from the others by the statistical analysis given below.

The second step for the evaluation of the light curve is to incorporate the neutrino oscillations, which take place as the Mikheyev-Smirnov-Wolfenstein (MSW) effect during the propagations through the stellar envelope and the Earth. Since the reactions at the detector depend on the neutrino flavor, a proper treatment of the neutrino oscillations is mandatory. In this paper, we ignore the Earth effect, since the event number will not be affected very much \cite{self08b}. As for the undetermined parameters in the neutrino oscillation, namely the mixing angle, $\theta_{13}$, and the mass hierarchy, we choose two limiting cases: the normal mass hierarchy with $\sin^2\theta_{13}=10^{-8}$ and the inverted mass hierarchy with $\sin^2\theta_{13}=10^{-2}$. For the detailed dependences on the choice of these parameters, we refer readers to Ref.~\cite{self08b}.

The last part of the procedure is an evaluation of event numbers at the detector. In this paper, we take SK from currently operating neutrino detectors. The neutrino reactions in the detector that we take into account are as follows:
\begin{eqnarray}
\label{ibd}
\bar\nu_e + p & \longrightarrow & e^+ + n, \\
\label{es}
\nu + e & \longrightarrow & \nu + e, \\
\label{ibdf}
\nu_e + \,^{16}\mathrm{O} & \longrightarrow & e + \,^{16}\mathrm{F}, \\
\label{ibdn}
\bar\nu_e + \,^{16}\mathrm{O} & \longrightarrow & e^+ + \,^{16}\mathrm{N}.
\end{eqnarray}
The first reaction gives a dominant contribution to the event number and we take its cross section from Ref.~\cite{strumia03}. The second reaction occurs for all flavors of neutrino but with different cross sections, which are taken from Ref.~\cite{totsuka92}. The cross sections for the others are adopted from Ref.~\cite{haxton87}. We assume that the fiducial volume is 22.5~kton and the trigger efficiency is 100\% at 4.5~MeV and 0\% at 2.9~MeV, which are the values at the end of SuperKamiokande I \cite{hosaka06}. The energy resolution was 14.2\% for $E_e=10$~MeV at that time \cite{hosaka06} and roughly proportional to $\sqrt{E_e}$ \cite{luna01}, where $E_e$ is the kinetic energy of scattered electrons and positrons. We choose the width of energy bin to be 1~MeV in this study. The event numbers for the progenitor at the distance of 10~kpc are listed in Table~\ref{modellist} for eight different combinations of EOS and mixing parameters.

\begin{table}[t]
\caption{Event numbers for the progenitor at the distance of 10~kpc. $N^\mathrm{10 kpc}_\mathrm{all}$ and $N^\mathrm{10 kpc}_\mathrm{0.5\,s}$ denote the event numbers until the end of simulation and up to 0.5~sec after bounce, respectively.}
\begin{center}
\setlength{\tabcolsep}{8pt}
\scalebox{0.9}{
\begin{tabular}{cccc}
\hline\hline
 EOS & Mixing parameter & $N^\mathrm{10 kpc}_\mathrm{all}$ & $N^\mathrm{10 kpc}_\mathrm{0.5\,s}$ \\ \hline
 LS180 & Normal \& $\sin^2\theta_{13}=10^{-8}$ & 16086 & 12543 \\
 LS220 & $\prime\prime$ & 25978 & 11970 \\
 Hyperon & $\prime\prime$ & 16490 & 10120 \\
 Shen & $\prime\prime$ & 49513 & 9745 \\ \hline
 LS180 & Inverted \& $\sin^2\theta_{13}=10^{-2}$ & 12136 & 9169 \\
 LS220 & $\prime\prime$ & 23656 & 8820 \\
 Hyperon & $\prime\prime$ & 9952 & 6579 \\
 Shen & $\prime\prime$ & 30992 & 6208 \\
\hline\hline
\end{tabular}
}
\label{modellist}
\end{center}
\end{table}

\subsection{Statistical analysis}

In order to see if two different hadronic EOS's can be distinguished by the neutrino observations, we take the following strategy. We first generate the ``observational data'' by Monte Carlo simulations based on the light curve obtained above for one EOS. We then take the light curve for another EOS as the ``theoretical'' model and employ the KS test to judge if the difference is statistically significant. In this study, we adopt Hyperon-EOS for the former and LS180- and LS220-EOS's for the latter.

In our analysis, we utilize the normalized time profiles of the cumulative event numbers for the first 0.5~sec of detections so that we could ignore the neutrino emissions after the black hole formation and the uncertainties of the distance measurement to the progenitor. In this study, we consider two cases, in which the total event numbers up to 0.5~sec, $N_\mathrm{0.5\,s}$, are 10000 and 400. The former roughly corresponds to the source located at the Galactic center while the latter represents an event in the Large Magellanic Cloud (LMC) or Small Magellanic Cloud (SMC).

In the KS test, the so-called KS measure, $D_\mathrm{KS}$, is defined as the maximum difference \cite{kendall79} between the two data:
\begin{equation}
D_\mathrm{KS} = \max_{t\le0.5\,\mathrm{s}} \bigl| f_\mathrm{theor.}(t) - f_\mathrm{obs.}(t) \bigr|,
\end{equation}
where $f_\mathrm{theor.}(t)$ and $f_\mathrm{obs.}(t)$ are the theoretical and observational time profiles of neutrino events, respectively. As mentioned already, they are normalized as $f_\mathrm{theor.}(0.5\,\mathrm{s})=f_\mathrm{obs.}(0.5\,\mathrm{s})=1$ in our analysis. If $D_\mathrm{KS}>0.01622$ (0.0811) for $N_\mathrm{0.5\,s}=10000$ (400), the theoretical model is rejected at the confidence level of 99\%.

We also employ time-shifted observational data, since we can not know the onset of neutrino detections precisely. In this case, the KS measure becomes a function of the time shift, $t_\mathrm{shift}$, as
\begin{equation}
D_\mathrm{KS}(t_\mathrm{shift}) = \max_{t\le0.5\,\mathrm{s}} \bigl| f_\mathrm{theor.}(t) - f_\mathrm{obs.}(t-t_\mathrm{shift}) \bigr|,
\end{equation}
where the time profiles are normalized as $f_\mathrm{theor.}(0.5\,\mathrm{s})=f_\mathrm{obs.} (0.5\,\mathrm{s}-t_\mathrm{shift})=1$. If the minimum value of the KS measure with respect to the time shift satisfies the above criteria, i.e., $\displaystyle \min_{t_\mathrm{shift}} D_\mathrm{KS} (t_\mathrm{shift})>0.01622$ (0.0811) for $N_\mathrm{0.5\,s}=10000$ (400), we can conclude at 99\% C.L. that the two data are distinct from each other. 

\section{Results} \label{result}

\subsection{Analysis without time-shift}

\begin{figure}
\begin{center}
\includegraphics{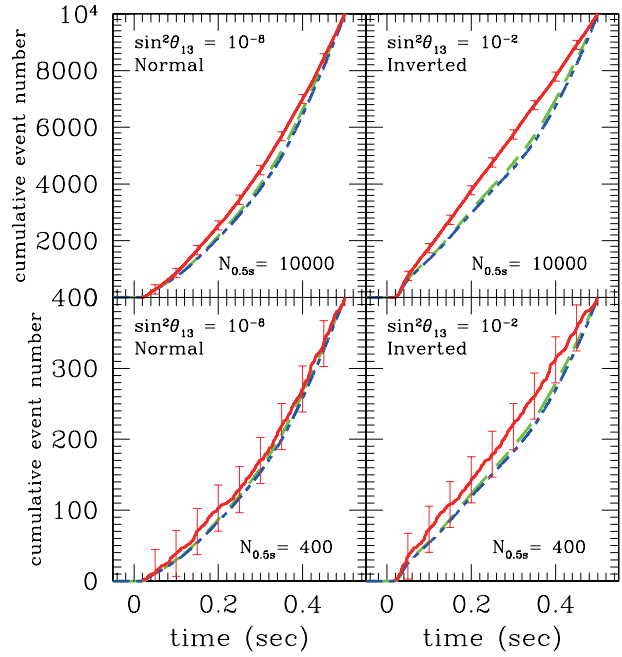}
\caption{Un-normalized time profiles of cumulative event numbers for two mixing parameter sets and two total event numbers. The observational data are shown by solid lines whereas the theoretical estimations are displayed by dash-dotted lines for LS180-EOS and dashed lines for LS220-EOS. The error bars are corresponding to 99\% C.L. in the KS test.}
\label{modedeps}
\end{center}
\end{figure}

We first show the results of the analysis without the time-shift. In Fig.~\ref{modedeps}, the un-normalized time profiles of cumulative event numbers are compared between a MC data (``observational data'') randomly picked up from 100000 realizations for Hyperon-EOS and the ``theoretical'' estimations for LS180-EOS and LS220-EOS. Since the growth rates of neutrino luminosity and mean energy are larger for LS-EOS's than for Hyperon-EOS (see Fig.~\ref{sumires}), the lines for the former are lower than for the latter. The vertical bars are corresponding to $\pm0.01622$ (0.0811) in the normalized profiles for $N_\mathrm{0.5\,s}=10000$ (400) and the theoretical models outside these error bars are rejected by the KS test at 99\% C.L. It is clear from the figure that for $N_\mathrm{0.5\,s}=10000$ or the Galactic source, we can easily distinguish Hyperon-EOS from LS-EOS's. In fact, this is true not only for this particular data but also for all 100000 MC realizations in each set of the mixing parameters. Thus we can conclude that Hyperon-EOS and LS-EOS's employed in this study are distinguishable for Galactic events.

The claim depends on the mixing parameters for events in LMC or SMC, that is, for $N_\mathrm{0.5\,s}=400$ as can be inferred from Fig.~\ref{modedeps}. In fact, for the normal mass hierarchy with $\sin^2\theta_{13}=10^{-8}$, Hyperon-EOS is not distinguishable from LS220-EOS (LS180-EOS) for more than 65\% (40\%) of 100000 MC realizations. In the case of the inverted mass hierarchy with $\sin^2\theta_{13}=10^{-2}$, however, the distinction fails 5218 (512) times among 100000 MC realizations for LS220-EOS (LS180-EOS). The reason why the latter case of the mixing parameters is easier is as follows. The EOS dependence is stronger for muon-type and tau-type neutrinos and their anti-neutrinos than for electron-type counterparts (see Fig.~3 of Ref.~\cite{sumi09}). A portion of $\bar{\nu}_{\mu}$ and $\bar{\nu}_{\tau}$ is converted to $\bar{\nu}_{e}$, the dominant contributor to the event number via reaction (\ref{ibd}). While 16\% of $\bar{\nu}_{\mu}$ and $\bar{\nu}_{\tau}$ become $\bar{\nu}_{e}$ for the normal mass hierarchy with $\sin^2\theta_{13}=10^{-8}$, a half of $\bar{\nu}_{\mu}$ and $\bar{\nu}_{\tau}$ are converted to $\bar{\nu}_{e}$ for the inverted mass hierarchy with $\sin^2\theta_{13}=10^{-2}$, thus yielding the observed EOS dependence of the event numbers.

\subsection{Analysis with time-shift}

\begin{figure}
\begin{center}
\includegraphics{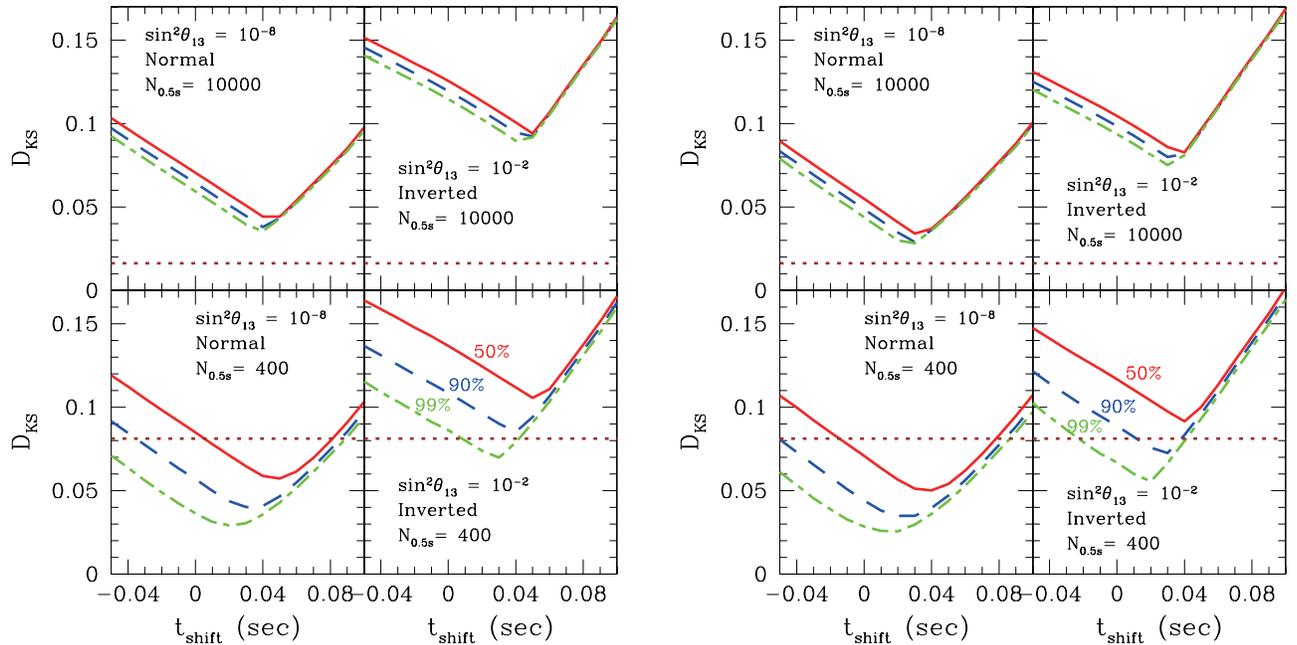}
\caption{The KS measures, $D_\mathrm{KS}$, as a function of the time-shift, $t_\mathrm{shift}$, for LS180-EOS (left panel) and LS220-EOS (right panel). 50\%, 90\% and 99\% of $D_\mathrm{KS}$'s lie above the solid, dashed and dot-dashed lines in 100000 MC realizations, respectively. The horizontal dotted lines represent the lower limit to reject the model by KS test.}
\label{tshift}
\end{center}
\end{figure}

Next, we demonstrate that the conclusion is unchanged if the time-shift is included in the analysis. In Fig.~\ref{tshift}, the distributions of the KS measures for 100000 MC realizations are shown as a function of the time-shift, $t_\mathrm{shift}$. LS180-EOS (LS220-EOS) is employed to obtain the ``theoretical'' model in the left (right) panel. For Galactic sources, that is, in the case of $N_\mathrm{0.5\,s}=10000$, all of 100000 MC realizations have $D_\mathrm{KS}$ larger than 0.01622, the lower limit for the rejection by the KS test, for any value of the time-shift in the case of the normal mass hierarchy with $\sin^2\theta_{13}=10^{-8}$. This is all the more true for the inverted mass hierarchy with $\sin^2\theta_{13}=10^{-2}$.

For $N_\mathrm{0.5\,s}=400$, however, the above conclusion is somewhat compromised. In the case of the normal mass hierarchy with $\sin^2\theta_{13}=10^{-8}$, for example, Hyperon-EOS and LS180-EOS are not distinguishable in more than 90\% of 100000 MC realizations when $t_\mathrm{shift}=0.06$~sec is assumed. For the inverted mass hierarchy with $\sin^2\theta_{13}=10^{-2}$, on the other hand, more than 75\% (90\%) of MC realizations have a KS measure larger than the critical value for LS220-EOS (LS180-EOS). In conclusion, for the event in LMC or SMC, the successful distinction of Hyperon- and LS-EOS's depends on the neutrino mixing parameters; the inverted mass hierarchy with $\sin^2\theta_{13}=10^{-2}$ is more advantageous.

\section{Discussion} \label{disc}

Certainly, there are some issues remaining unaddressed. First of all, investigating other EOS's constructed by different manners is important future work. In particular, hyperon mixing in neutron star matter has been studied by the microscopic approach \cite{refa, refa2, refa3} and it is desired to be applied to EOS tables including finite temperature. If neutrinos have masses of $\sim$~eV, the arrival time depends on the neutrino energy and the difference becomes $\sim \!\! 0.01$~sec between the neutrinos with an energy of $E \sim 10$~MeV and those with $\gtrsim \!\! 30$~MeV \cite{beacom00} for Galactic sources. Recently, the core collapse simulations for the same progenitor model are done utilizing a spherically symmetric $\nu$-radiation-hydrodynamics code by another grope \cite{fischer08}. While they conclude that their results are in qualitative agreement with our study, there are some quantitative differences. For instance, in the case with LS180-EOS, they found that the interval time between the bounce and black hole formation is 0.436~sec, which is $\sim \!\! 20$\% shorter than our result. Further numerical investigation is important.

Multi-dimensional hydrodynamical effects such as rotations and magnetic fields, which are not taken into account in our model, may also affect the signals. In the context of supernovae explosions, Ref.~\cite{marek09} compared the neutrino signal by an one-dimensional simulation with those by two-dimensional simulations (with and without rotation) for the collapse of a $15M_\odot$ star adopting LS180-EOS (see Figs.~1 and 14 of this paper). According to this paper, when the initial rotation of progenitor core is a constant angular frequency of 0.5~rad~s$^{-1}$, the stellar rotation does not affect the luminosity and mean energy of the emitted neutrinos very much. Incidentally, the neutrino emission after the collapse to a black hole is estimated to be negligible for low-angular momentum cases \cite{fryer09}.

Uncertainties with progenitor structures are another concern. While a certain class of progenitors provides similarly short neutrino burst, it has been shown that the density profile of outer layer may affect the duration \cite{sumi08}. Especially, in the case with lower matter density outside the core \cite{woosley02}, the interval time between the bounce and black hole formation becomes longer as 1.477~sec even for the model with $40M_\odot$ and LS180-EOS \cite{fischer08}. However, if a neutrino event is actually detected, we will be able to determine the direction of the progenitor to some extent by the neutrino detection itself \cite{ando02} and the progenitor is highly likely to be identified or at least constrained by earlier records of optical observations as in the case of SN1987A. Then we can study the progenitor dependence much more efficiently.

As discussed in Refs.~\cite{kocha08, self08b}, the event rate of the black-hole-forming failed supernovae is estimated to be somewhat low $\sim \!\! 0.008$/yr. This problem may be circumvented, however, by deploying a large detector such as Deep-TITAND with a fiducial volume of 5~Mton \cite{kist08} proposed currently. With this large facility, the event rate goes up to $\sim \!\! 0.02$/yr since we should be able to detect $\sim \!\! 400$ neutrinos for the first 0.5~sec of a failed supernova in galaxies as far away as the Andromeda galaxy (M31) at 780~kpc from us. This is large enough to distinguish the EOS's we studied in this letter for the inverted mass hierarchy with $\sin^2\theta_{13}=10^{-2}$. Note that the cumulative core-collapse supernova rate within this range is 10\% of that within 10~Mpc \cite{ando05}, where a regular monitoring of $\sim \!\! 10^6$ supergiants is proposed in Ref.~\cite{kocha08}.

\section{Summary} \label{sum}

This study is the first ever serious self-contained attempt to assess the neutrino signals from black-hole-forming failed supernovae, which would be observed by the currently operating terrestrial detector. Based on our detailed numerical simulations, we have evaluated the event number of neutrinos emitted from black-hole-forming failed supernova for some EOS's of nuclear matter. Ambiguities on the neutrino mixing parameters and the onset of the neutrino emission have also been taken into account for the evaluation. Assuming the same progenitor model and non-rotation, we have shown that we will be able to constrain the EOS of nuclear matter not only from the duration time of neutrino emission but also from the time variations of neutrino event number for the progenitor in our Galaxy. Moreover, in the case of the inverted mass hierarchy with $\sin^2\theta_{13}=10^{-2}$, the constraint is favorable even for the progenitor in LMC or SMC. The positive results presented here clearly indicate the importance of further investigations of the hadronic EOS at supra-nuclear densities based on better formulations and encourage in particular those engaged in the study of the EOS of hadronic matter to prepare their latest results in a form available for astrophysical simulations. We hope this paper will advance such collaborations further.

\begin{acknowledgments}
We are grateful to Akira Ohnishi and Chikako Ishizuka for fruitful discussions. In this work, numerical computations were partially performed on the supercomputers at Research Center for Nuclear Physics (RCNP) in Osaka University, Center for Computational Astrophysics (CfCA) in the National Astronomical Observatory of Japan (NAOJ), Yukawa Institute for Theoretical Physics (YITP) in Kyoto University, Japan Atomic Energy Agency (JAEA) and High Energy Accelerator Research Organization (KEK). This work was partially supported by Research Fellowship for Young Scientists from the Japan Society for Promotion of Science (JSPS) through 18-510 and 21-1189, and Grants-in-Aid for the Scientific Research from the Ministry of Education, Culture, Sports, Science and Technology (MEXT) in Japan through 17540267, 18540291, 18540295, 19104006, 19540252, 20105004 and 21540281.
\end{acknowledgments}

\bibliographystyle{apsrev} 
\bibliography{apssamp}% Produces the bibliography via BibTeX.

\begin{thebibliography}{45}
\expandafter\ifx\csname natexlab\endcsname\relax\def\natexlab#1{#1}\fi
\expandafter\ifx\csname bibnamefont\endcsname\relax
  \def\bibnamefont#1{#1}\fi
\expandafter\ifx\csname bibfnamefont\endcsname\relax
  \def\bibfnamefont#1{#1}\fi
\expandafter\ifx\csname citenamefont\endcsname\relax
  \def\citenamefont#1{#1}\fi
\expandafter\ifx\csname url\endcsname\relax
  \def\url#1{\texttt{#1}}\fi
\expandafter\ifx\csname urlprefix\endcsname\relax\def\urlprefix{URL }\fi
\providecommand{\bibinfo}[2]{#2}
\providecommand{\eprint}[2][]{\url{#2}}


%\bibitem[{\citenamefont{Sumiyoshi et~al.}(2009)}]{sumi09}
%\bibinfo{author}{\bibfnamefont{K.} \bibnamefont{Sumiyoshi}} \bibnamefont{{\it et~al.}},
%  \bibinfo{journal}{Astrophys.\ J.} \textbf{\bibinfo{volume}{690}},
%  \bibinfo{pages}{L43} (\bibinfo{year}{2009}).

\bibitem[{\citenamefont{Sumiyoshi et~al.}(2009)\citenamefont{Sumiyoshi, Ishizuka, Ohnishi, Yamada, and Suzuki}}]{sumi09}
\bibinfo{author}{\bibfnamefont{K.}~\bibnamefont{Sumiyoshi}},
  \bibinfo{author}{\bibfnamefont{C.}~\bibnamefont{Ishizuka}},
  \bibinfo{author}{\bibfnamefont{A.}~\bibnamefont{Ohnishi}},
  \bibinfo{author}{\bibfnamefont{S.}~\bibnamefont{Yamada}}, \bibnamefont{and}
  \bibinfo{author}{\bibfnamefont{H.}~\bibnamefont{Suzuki}},
  \bibinfo{journal}{Astrophys.\ J.} \textbf{\bibinfo{volume}{690}},
  \bibinfo{pages}{L43} (\bibinfo{year}{2009}).

\bibitem[{\citenamefont{Nakazato et~al.}(2008a)\citenamefont{Nakazato, Sumiyoshi, and Yamada}}]{self08a}
\bibinfo{author}{\bibfnamefont{K.}~\bibnamefont{Nakazato}},
  \bibinfo{author}{\bibfnamefont{K.}~\bibnamefont{Sumiyoshi}},
  \bibnamefont{and} \bibinfo{author}{\bibfnamefont{S.}~\bibnamefont{Yamada}},
  \bibinfo{journal}{Phys.\ Rev.\ D} \textbf{\bibinfo{volume}{77}},
  \bibinfo{pages}{103006} (\bibinfo{year}{2008a}).

\bibitem[{\citenamefont{Prestwich et~al.}(2007)}]{prest07}
\bibinfo{author}{\bibfnamefont{A.~H.}~\bibnamefont{Prestwich}} \bibnamefont{{\it et~al.}},
  \bibinfo{journal}{Astrophys.\ J.} \textbf{\bibinfo{volume}{669}},
  \bibinfo{pages}{L21} (\bibinfo{year}{2007}).

\bibitem[{\citenamefont{Kochanek et~al.}(2008)}]{kocha08}
\bibinfo{author}{\bibfnamefont{C.~S.}~\bibnamefont{Kochanek}} \bibnamefont{{\it et~al.}},
  \bibinfo{journal}{Astrophys.\ J.} \textbf{\bibinfo{volume}{684}},
  \bibinfo{pages}{1336} (\bibinfo{year}{2008}).

\bibitem[{\citenamefont{Sumiyoshi et~al.}(2006)\citenamefont{Sumiyoshi, Yamada, Suzuki, and Chiba}}]{sumi06}
\bibinfo{author}{\bibfnamefont{K.}~\bibnamefont{Sumiyoshi}},
  \bibinfo{author}{\bibfnamefont{S.}~\bibnamefont{Yamada}},
  \bibinfo{author}{\bibfnamefont{H.}~\bibnamefont{Suzuki}}, \bibnamefont{and}
  \bibinfo{author}{\bibfnamefont{S.}~\bibnamefont{Chiba}},
  \bibinfo{journal}{Phys.\ Rev.\ Lett.} \textbf{\bibinfo{volume}{97}},
  \bibinfo{pages}{091101} (\bibinfo{year}{2006}).

\bibitem[{\citenamefont{Sumiyoshi et~al.}(2006)\citenamefont{Sumiyoshi, Yamada, and Suzuki}}]{sumi07}
\bibinfo{author}{\bibfnamefont{K.}~\bibnamefont{Sumiyoshi}},
  \bibinfo{author}{\bibfnamefont{S.}~\bibnamefont{Yamada}}, \bibnamefont{and}
  \bibinfo{author}{\bibfnamefont{H.}~\bibnamefont{Suzuki}},
  \bibinfo{journal}{Astrophys.\ J.} \textbf{\bibinfo{volume}{667}},
  \bibinfo{pages}{382} (\bibinfo{year}{2007}).

\bibitem[{\citenamefont{Sumiyoshi et~al.}(2008)\citenamefont{Sumiyoshi, Yamada,
  and Suzuki}}]{sumi08}
\bibinfo{author}{\bibfnamefont{K.}~\bibnamefont{Sumiyoshi}},
  \bibinfo{author}{\bibfnamefont{S.}~\bibnamefont{Yamada}}, \bibnamefont{and}
  \bibinfo{author}{\bibfnamefont{H.}~\bibnamefont{Suzuki}},
  \bibinfo{journal}{Astrophys.\ J.} \textbf{\bibinfo{volume}{688}},
  \bibinfo{pages}{1176} (\bibinfo{year}{2008}).

\bibitem[{\citenamefont{Totani et~al.}(1998)\citenamefont{Totani, Sato, Dalhed,
  and Wilson}}]{totani98}
\bibinfo{author}{\bibfnamefont{T.}~\bibnamefont{Totani}},
  \bibinfo{author}{\bibfnamefont{K.}~\bibnamefont{Sato}},
  \bibinfo{author}{\bibfnamefont{H.~E.} \bibnamefont{Dalhed}},
  \bibnamefont{and} \bibinfo{author}{\bibfnamefont{J.~R.}
  \bibnamefont{Wilson}}, \bibinfo{journal}{Astrophys.\ J.}
  \textbf{\bibinfo{volume}{496}}, \bibinfo{pages}{216} (\bibinfo{year}{1998}).

\bibitem[{\citenamefont{Baumgarte et~al.}(1996)\citenamefont{Baumgarte, Janka, Keil, Shapiro, and Teukolsky}}]{refb}
\bibinfo{author}{\bibfnamefont{T.~W.} \bibnamefont{Baumgarte}},
  \bibinfo{author}{\bibfnamefont{H.-Th.}~\bibnamefont{Janka}},
  \bibinfo{author}{\bibfnamefont{W.}~\bibnamefont{Keil}},
  \bibinfo{author}{\bibfnamefont{S.~L.}~\bibnamefont{Shapiro}}, \bibnamefont{and}
  \bibinfo{author}{\bibfnamefont{S.~A.}~\bibnamefont{Teukolsky}},
  \bibinfo{journal}{Astrophys.\ J.} \textbf{\bibinfo{volume}{468}},
  \bibinfo{pages}{823} (\bibinfo{year}{1996}).

\bibitem{refb2}
\bibinfo{author}{\bibfnamefont{J.~A.} \bibnamefont{Pons}},
  \bibinfo{author}{\bibfnamefont{A.~W.}~\bibnamefont{Steiner}},
  \bibinfo{author}{\bibfnamefont{M.}~\bibnamefont{Prakash}}, \bibnamefont{and}
  \bibinfo{author}{\bibfnamefont{J.~M.}~\bibnamefont{Lattimer}},
  \bibinfo{journal}{Phys.\ Rev.\ Lett.} \textbf{\bibinfo{volume}{86}},
  \bibinfo{pages}{5223} (\bibinfo{year}{2001}).

\bibitem[{\citenamefont{Lunardini}(2009)}]{luna09}
\bibinfo{author}{\bibfnamefont{C.}~\bibnamefont{Lunardini}},
  \bibinfo{journal}{Phys.\ Rev.\ Lett.} \textbf{\bibinfo{volume}{102}},
  \bibinfo{pages}{231101} (\bibinfo{year}{2009}).

\bibitem[{\citenamefont{Nakazato et~al.}(2008b)\citenamefont{Nakazato, Sumiyoshi, Suzuki, and Yamada}}]{self08b}
\bibinfo{author}{\bibfnamefont{K.}~\bibnamefont{Nakazato}},
  \bibinfo{author}{\bibfnamefont{K.}~\bibnamefont{Sumiyoshi}},
  \bibinfo{author}{\bibfnamefont{H.}~\bibnamefont{Suzuki}}, \bibnamefont{and}
  \bibinfo{author}{\bibfnamefont{S.}~\bibnamefont{Yamada}},
  \bibinfo{journal}{Phys.\ Rev.\ D} \textbf{\bibinfo{volume}{78}},
  \bibinfo{pages}{083014} (\bibinfo{year}{2008b}).

\bibitem{yamada97}
\bibinfo{author}{\bibfnamefont{S.}~\bibnamefont{Yamada}},
  \bibinfo{journal}{Astrophys.\ J.} \textbf{\bibinfo{volume}{475}},
  \bibinfo{pages}{720} (\bibinfo{year}{1997}).

\bibitem{yamada99}
\bibinfo{author}{\bibfnamefont{S.}~\bibnamefont{Yamada}},
  \bibinfo{author}{\bibfnamefont{H.-Th}~\bibnamefont{Janka}}, \bibnamefont{and}
  \bibinfo{author}{\bibfnamefont{H.}~\bibnamefont{Suzuki}},
  \bibinfo{journal}{Astron.\ Astrophys.} \textbf{\bibinfo{volume}{344}},
  \bibinfo{pages}{533} (\bibinfo{year}{1999}).

\bibitem{sumi05}
\bibinfo{author}{\bibfnamefont{K.} \bibnamefont{Sumiyoshi}} \bibnamefont{{\it et~al.}},
  \bibinfo{journal}{Astrophys.\ J.} \textbf{\bibinfo{volume}{629}},
  \bibinfo{pages}{922} (\bibinfo{year}{2005}).

\bibitem[{\citenamefont{Nakazato et~al.}(2007)\citenamefont{Nakazato,
  Sumiyoshi, and Yamada}}]{self07}
\bibinfo{author}{\bibfnamefont{K.}~\bibnamefont{Nakazato}},
  \bibinfo{author}{\bibfnamefont{K.}~\bibnamefont{Sumiyoshi}},
  \bibnamefont{and} \bibinfo{author}{\bibfnamefont{S.}~\bibnamefont{Yamada}},
  \bibinfo{journal}{Astrophys.\ J.} \textbf{\bibinfo{volume}{666}},
  \bibinfo{pages}{1140} (\bibinfo{year}{2007}).

\bibitem[{\citenamefont{Woosley and Weaver}(1995)}]{woosley95}
\bibinfo{author}{\bibfnamefont{S.~E.} \bibnamefont{Woosley}} \bibnamefont{and}
  \bibinfo{author}{\bibfnamefont{T.}~\bibnamefont{Weaver}},
  \bibinfo{journal}{Astrophys.\ J.\ Suppl.} \textbf{\bibinfo{volume}{101}},
  \bibinfo{pages}{181} (\bibinfo{year}{1995}).

%\bibitem[{\citenamefont{Ishizuka et~al.}(2008)\citenamefont{Ishizuka et~al.}}]{takochu}
%\bibinfo{author}{\bibfnamefont{C.}~\bibnamefont{Ishizuka}} \bibnamefont{{\it et~al.}},
%  \bibinfo{journal}{J.\ Phys.\ G} \textbf{\bibinfo{volume}{35}},
%  \bibinfo{pages}{085201} (\bibinfo{year}{2008}).

\bibitem[{\citenamefont{Ishizuka et~al.}(2008)\citenamefont{Ishizuka, Ohnishi, Tsubakihara, Sumiyoshi, and Yamada}}]{takochu}
\bibinfo{author}{\bibfnamefont{C.}~\bibnamefont{Ishizuka}},
  \bibinfo{author}{\bibfnamefont{A.}~\bibnamefont{Ohnishi}},
  \bibinfo{author}{\bibfnamefont{K.}~\bibnamefont{Tsubakihara}},
  \bibinfo{author}{\bibfnamefont{K.}~\bibnamefont{Sumiyoshi}},
  \bibnamefont{and} \bibinfo{author}{\bibfnamefont{S.}~\bibnamefont{Yamada}},
  \bibinfo{journal}{J.\ Phys.\ G} \textbf{\bibinfo{volume}{35}},
  \bibinfo{pages}{085201} (\bibinfo{year}{2008}).

\bibitem[{\citenamefont{Lattimer and Swesty}(1991)}]{lati91}
\bibinfo{author}{\bibfnamefont{J.~M.} \bibnamefont{Lattimer}} \bibnamefont{and}
  \bibinfo{author}{\bibfnamefont{F.~D.} \bibnamefont{Swesty}},
  \bibinfo{journal}{Nucl.\ Phys.\ A} \textbf{\bibinfo{volume}{535}},
  \bibinfo{pages}{331} (\bibinfo{year}{1991}).

\bibitem[{\citenamefont{Shen et~al.}(1998a)\citenamefont{Shen, Toki, Oyamatsu,
  and Sumiyoshi}}]{shen98a}
\bibinfo{author}{\bibfnamefont{H.}~\bibnamefont{Shen}},
  \bibinfo{author}{\bibfnamefont{H.}~\bibnamefont{Toki}},
  \bibinfo{author}{\bibfnamefont{K.}~\bibnamefont{Oyamatsu}}, \bibnamefont{and}
  \bibinfo{author}{\bibfnamefont{K.}~\bibnamefont{Sumiyoshi}},
  \bibinfo{journal}{Nucl.\ Phys.\ A} \textbf{\bibinfo{volume}{637}},
  \bibinfo{pages}{435} (\bibinfo{year}{1998a}).

\bibitem[{\citenamefont{Shen et~al.}(1998b)\citenamefont{Shen, Toki, Oyamatsu,
  and Sumiyoshi}}]{shen98b}
\bibinfo{author}{\bibfnamefont{H.}~\bibnamefont{Shen}},
  \bibinfo{author}{\bibfnamefont{H.}~\bibnamefont{Toki}},
  \bibinfo{author}{\bibfnamefont{K.}~\bibnamefont{Oyamatsu}}, \bibnamefont{and}
  \bibinfo{author}{\bibfnamefont{K.}~\bibnamefont{Sumiyoshi}},
  \bibinfo{journal}{Prog.\ Theor.\ Phys.} \textbf{\bibinfo{volume}{100}},
  \bibinfo{pages}{1013} (\bibinfo{year}{1998b}).
%
%\bibitem[{\citenamefont{Shen et~al.}(1998a)\citenamefont{Shen, Toki, Oyamatsu,
%  and Sumiyoshi}}]{shen98a}
%\bibinfo{author}{\bibfnamefont{H.}~\bibnamefont{Shen}},
%  \bibinfo{author}{\bibfnamefont{H.}~\bibnamefont{Toki}},
%  \bibinfo{author}{\bibfnamefont{K.}~\bibnamefont{Oyamatsu}}, \bibnamefont{and}
%  \bibinfo{author}{\bibfnamefont{K.}~\bibnamefont{Sumiyoshi}},
%  \bibinfo{journal}{Nucl.\ Phys.\ A} \textbf{\bibinfo{volume}{637}},
%  \bibinfo{pages}{435} (\bibinfo{year}{1998a}),
%
%\bibitem[{\citenamefont{Shen et~al.}(1998b)\citenamefont{Shen, Toki, Oyamatsu,
%  and Sumiyoshi}}]{shen98b}
%\bibinfo{author}{\bibfnamefont{H.}~\bibnamefont{Shen}},
%  \bibinfo{author}{\bibfnamefont{H.}~\bibnamefont{Toki}},
%  \bibinfo{author}{\bibfnamefont{K.}~\bibnamefont{Oyamatsu}}, \bibnamefont{and}
%  \bibinfo{author}{\bibfnamefont{K.}~\bibnamefont{Sumiyoshi}},
%  \bibinfo{journal}{Prog.\ Theor.\ Phys.} \textbf{\bibinfo{volume}{100}},
%  \bibinfo{pages}{1013} (\bibinfo{year}{1998b}).

%\bibitem[{\citenamefont{Yamada}(1997)}]{yamada97}
%\bibinfo{author}{\bibfnamefont{S.}~\bibnamefont{Yamada}},
%  \bibinfo{journal}{Astrophys.\ J.} \textbf{\bibinfo{volume}{475}},
%  \bibinfo{pages}{720} (\bibinfo{year}{1997}).

\bibitem[{\citenamefont{Strumia and Vissani}(2003)}]{strumia03}
\bibinfo{author}{\bibfnamefont{A.}~\bibnamefont{Strumia}} \bibnamefont{and}
  \bibinfo{author}{\bibfnamefont{F.}~\bibnamefont{Vissani}},
  \bibinfo{journal}{Phys.\ Lett.\ B} \textbf{\bibinfo{volume}{564}},
  \bibinfo{pages}{42} (\bibinfo{year}{2003}).

\bibitem[{\citenamefont{Totsuka}(1992)}]{totsuka92}
\bibinfo{author}{\bibfnamefont{Y.}~\bibnamefont{Totsuka}},
  \bibinfo{journal}{Rep.\ Prog.\ Phys.} \textbf{\bibinfo{volume}{55}},
  \bibinfo{pages}{377} (\bibinfo{year}{1992}).

\bibitem[{\citenamefont{Haxton}(1987)}]{haxton87}
\bibinfo{author}{\bibfnamefont{W.~C.} \bibnamefont{Haxton}},
  \bibinfo{journal}{Phys.\ Rev.\ D} \textbf{\bibinfo{volume}{36}},
  \bibinfo{pages}{2283} (\bibinfo{year}{1987}).

\bibitem[{\citenamefont{Hosaka et~al.}(2006)}]{hosaka06}
\bibinfo{author}{\bibfnamefont{J.}~\bibnamefont{Hosaka}} \bibnamefont{{\it et~al.}},
  \bibinfo{journal}{Phys.\ Rev.\ D} \textbf{\bibinfo{volume}{73}},
  \bibinfo{pages}{112001} (\bibinfo{year}{2006}).

\bibitem[{\citenamefont{Lunardini and Smirnov}(2001)}]{luna01}
\bibinfo{author}{\bibfnamefont{C.}~\bibnamefont{Lunardini}} \bibnamefont{and}
  \bibinfo{author}{\bibfnamefont{A.~Y.} \bibnamefont{Smirnov}},
  \bibinfo{journal}{Nucl.\ Phys.\ B} \textbf{\bibinfo{volume}{616}},
  \bibinfo{pages}{307} (\bibinfo{year}{2001}).

\bibitem[{\citenamefont{Kendall and Stuart}(1979)}]{kendall79}
\bibinfo{author}{\bibfnamefont{M.~G.} \bibnamefont{Kendall}} \bibnamefont{and}
  \bibinfo{author}{\bibfnamefont{A.}~\bibnamefont{Stuart}},
  \emph{\bibinfo{title}{The Advanced Theory of Statistics}}
  (\bibinfo{publisher}{London: Griffin}, \bibinfo{year}{1979}).

\bibitem{refa}
%\bibinfo{author}{\bibfnamefont{M.} \bibnamefont{Baldo}} \bibnamefont{{\it et~al.}},
%  \bibinfo{journal}{Phys.\ Rev.\ C} \textbf{\bibinfo{volume}{58}},
%  \bibinfo{pages}{3688} (\bibinfo{year}{1998}),
%\bibinfo{author}{\bibfnamefont{I.} \bibnamefont{Vida\~{n}a}} \bibnamefont{{\it et~al.}},
%  \bibinfo{journal}{Phys.\ Rev.\ C} \textbf{\bibinfo{volume}{61}},
%  \bibinfo{pages}{025802} (\bibinfo{year}{2000}),
\bibinfo{author}{\bibfnamefont{M.} \bibnamefont{Baldo}},
  \bibinfo{author}{\bibfnamefont{G.~F.}~\bibnamefont{Burgio}},
  \bibnamefont{and} \bibinfo{author}{\bibfnamefont{H.-J.}~\bibnamefont{Schulze}},
  \bibinfo{journal}{Phys.\ Rev.\ C} \textbf{\bibinfo{volume}{61}},
  \bibinfo{pages}{055801} (\bibinfo{year}{2000}).

\bibitem{refa2}
\bibinfo{author}{\bibfnamefont{S.} \bibnamefont{Nishizaki}},
  \bibinfo{author}{\bibfnamefont{Y.}~\bibnamefont{Yamamoto}},
  \bibnamefont{and} \bibinfo{author}{\bibfnamefont{T.}~\bibnamefont{Takatsuka}},
  \bibinfo{journal}{Prog.\ Theor.\ Phys.} \textbf{\bibinfo{volume}{108}},
  \bibinfo{pages}{703} (\bibinfo{year}{2002}).

\bibitem{refa3}
\bibinfo{author}{\bibfnamefont{H.-J.} \bibnamefont{Schulze}},
  \bibinfo{author}{\bibfnamefont{A.}~\bibnamefont{Polls}},
  \bibinfo{author}{\bibfnamefont{A.}~\bibnamefont{Ramos}},
  \bibnamefont{and} \bibinfo{author}{\bibfnamefont{I.}~\bibnamefont{Vida\~{n}a}},
  \bibinfo{journal}{Phys.\ Rev.\ C} \textbf{\bibinfo{volume}{73}},
  \bibinfo{pages}{058801} (\bibinfo{year}{2006}).

\bibitem[{\citenamefont{Beacom et~al.}(2001)\citenamefont{Beacom, Boyd, and
  Mezzacappa}}]{beacom00}
\bibinfo{author}{\bibfnamefont{J.~F.} \bibnamefont{Beacom}},
  \bibinfo{author}{\bibfnamefont{R.~N.} \bibnamefont{Boyd}}, \bibnamefont{and}
  \bibinfo{author}{\bibfnamefont{A.}~\bibnamefont{Mezzacappa}},
  \bibinfo{journal}{Phys.\ Rev.\ D} \textbf{\bibinfo{volume}{63}},
  \bibinfo{pages}{073011} (\bibinfo{year}{2001}).

%\bibitem[{\citenamefont{Gonzalez-Garcia and Maltoni}(2008)}]{gongar08}
%\bibinfo{author}{\bibfnamefont{M.~C.} \bibnamefont{Gonzalez-Garcia}}
%  \bibnamefont{and} \bibinfo{author}{\bibfnamefont{M.}~\bibnamefont{Maltoni}},
%  \bibinfo{journal}{Phys.\ Rept.} \textbf{\bibinfo{volume}{460}},
%  \bibinfo{pages}{1} (\bibinfo{year}{2008}).

\bibitem[{\citenamefont{Fischer et~al.}(2009)}]{fischer08}
\bibinfo{author}{\bibfnamefont{T.}~\bibnamefont{Fischer}},
  \bibinfo{author}{\bibfnamefont{S.~C.}~\bibnamefont{Whitehouse}},
  \bibinfo{author}{\bibfnamefont{A.}~\bibnamefont{Mezzacappa}},
  \bibinfo{author}{\bibfnamefont{F.-K.}~\bibnamefont{Thielemann}}, \bibnamefont{and}
  \bibinfo{author}{\bibfnamefont{M.}~\bibnamefont{Liebend\"{o}rfer}},
  \bibinfo{journal}{Astron.\ Astrophys.} \textbf{\bibinfo{volume}{499}},
  \bibinfo{pages}{1} (\bibinfo{year}{2009}).

\bibitem[{\citenamefont{Marek and Janka}(2009)}]{marek09}
\bibinfo{author}{\bibfnamefont{A.} \bibnamefont{Marek}}, \bibnamefont{and}
  \bibinfo{author}{\bibfnamefont{H.-Th.}~\bibnamefont{Janka}},
  \bibinfo{journal}{Astrophys.\ J.} \textbf{\bibinfo{volume}{694}},
  \bibinfo{pages}{664} (\bibinfo{year}{2009}).

\bibitem{fryer09}
\bibinfo{author}{\bibfnamefont{C.~L.}~\bibnamefont{Fryer}},
  \bibinfo{journal}{Astrophys.\ J.} \textbf{\bibinfo{volume}{699}},
  \bibinfo{pages}{409} (\bibinfo{year}{2009}).

\bibitem[{\citenamefont{Woosley et~al.}(2002)}]{woosley02}
\bibinfo{author}{\bibfnamefont{S.~E.} \bibnamefont{Woosley}},
  \bibinfo{author}{\bibfnamefont{A.}~\bibnamefont{Heger}}, \bibnamefont{and}
  \bibinfo{author}{\bibfnamefont{T.}~\bibnamefont{Weaver}},
  \bibinfo{journal}{Rev.\ Mod.\ Phys.} \textbf{\bibinfo{volume}{74}},
  \bibinfo{pages}{1015} (\bibinfo{year}{2002}).

\bibitem[{\citenamefont{Ando and Sato}(2002)}]{ando02}
\bibinfo{author}{\bibfnamefont{S.}~\bibnamefont{Ando}}, \bibnamefont{and}
  \bibinfo{author}{\bibfnamefont{K.}~\bibnamefont{Sato}},
  \bibinfo{journal}{Prog.\ Theor.\ Phys.} \textbf{\bibinfo{volume}{107}},
  \bibinfo{pages}{957} (\bibinfo{year}{2002}).

\bibitem[{\citenamefont{Kistler et~al.}(2008)\citenamefont{Kistler, Y{\"u}ksel,
  Ando, Beacom, and Suzuki}}]{kist08}
\bibinfo{author}{\bibfnamefont{M.~D.} \bibnamefont{Kistler}},
  \bibinfo{author}{\bibfnamefont{H.}~\bibnamefont{Y{\"u}ksel}},
  \bibinfo{author}{\bibfnamefont{S.}~\bibnamefont{Ando}},
  \bibinfo{author}{\bibfnamefont{J.~F.} \bibnamefont{Beacom}},
  \bibnamefont{and} \bibinfo{author}{\bibfnamefont{Y.}~\bibnamefont{Suzuki}},
  \eprint{arXiv:0810.1959}.

\bibitem[{\citenamefont{Ando et~al.}(2005)\citenamefont{Ando, Beacom, and Y{\"u}ksel}}]{ando05}
\bibinfo{author}{\bibfnamefont{S.}~\bibnamefont{Ando}},
  \bibinfo{author}{\bibfnamefont{J.~F.}~\bibnamefont{Beacom}}, \bibnamefont{and}
  \bibinfo{author}{\bibfnamefont{H.}~\bibnamefont{Y{\"u}ksel}},
  \bibinfo{journal}{Phys.\ Rev.\ Lett.} \textbf{\bibinfo{volume}{95}},
  \bibinfo{pages}{171101} (\bibinfo{year}{2005}).


\end{thebibliography}

\end{document}